# Evaluation of the relativistic redshift in frequency standards at KRISS

Jisun Lee[1], Jay Hyoun Kwon[1,*], Chang Yong Park[2], Huidong Kim[2], In-Mook Choi[2], Jin Wan Chung[2], and Won-Kyu Lee[2,*]

[1] Department of Geoinformatics, University of Seoul, Seoul 02504, Republic of Korea

[2] Korea Research Institute of Standards and Science, Daejeon 34113, Republic of Korea

* E-mail: jkwon@uos.ac.kr and oneqlee@kriss.re.kr

## Abstract

Relativistic redshift correction should be accurately considered in frequency comparisons between frequency standards. In this study, we evaluated the relativistic redshift at Korea Research Institute of Standards and Science (KRISS) using three different methods, depending on whether the approach was traditional or modern or whether the geopotential model was global or local. The results of the three methods agreed well with one another, and the height of an Yb optical lattice clock (KRISS-Yb1) was determined to be 75.15 m with an uncertainty of 0.04 m with respect to the conventionally adopted equipotential surface $W_0^{CGPM}$, the value of which is defined to be 62 636 856.0 m$^2$s$^{-2}$. Accordingly, the relativistic redshift of KRISS-Yb1 was evaluated to be $8.193(4) \times 10^{-15}$. These data are applicable to the frequency standards at KRISS, one of which regularly participates in the calibration of the International Atomic Time (TAI).

## 1. Introduction

Currently, the SI unit of time (the second with the symbol s) is defined by the unperturbed ground-state hyperfine transition frequency of the [133]Cs atom [1], which is realized by Cs fountain clocks with uncertainties at low $10^{-16}$ level [2, 3]. However, in the past two decades, the performance of optical atomic clocks has rapidly progressed, surpassing that of Cs fountain clocks, and has achieved frequency uncertainties with a level of approximately $1 \times 10^{-18}$ [4-10]. Therefore, a redefinition of the second using optical clocks is being considered [11-14].

Because the frequencies of atomic clocks vary with gravity potential value according to general relativity, relativistic redshift frequency correction is indispensable for comparing atomic clocks or steering International Atomic Time (TAI). The relativistic redshift correction for a clock at a point $P$ at rest on the Earth's surface is approximately given by [15]

$$\frac{\Delta f}{f_P} \approx \frac{W_P - W_0}{c^2} = -\frac{gH}{c^2}, \tag{1}$$

where $f_P$ is the frequency measured at $P$, $\Delta f$ is its difference from that measured on the zero level surface, $(W_P - W_0)$ is the difference in gravity potential between $P$ and the zero level surface, $g$ is the gravity acceleration, $H$ is the height referenced to the zero level surface, and $c$ is the speed of light.

For this purpose, in 2018, the General Conference on Weights and Measures (CGPM) recommended when transforming the proper time of a clock to TAI, the relativistic redshift should be computed with respect to the conventionally adopted equipotential surface $W_0^{CGPM}$, the value of which is defined to be 62 636 856.0 m$^2$s$^{-2}$ [16], which conforms to the constant L$_G$ = 6.969 290 134 × 10$^{−10}$ to define the rate of terrestrial time (TT) [17].

Accordingly, several studies have been conducted to evaluate the relativistic redshift and to determine the heights of atomic clocks [15, 18-26]. The heights of atomic clocks were estimated with an uncertainty of 6 cm at NIST, Boulder, Colorado, USA [21], with an uncertainty of 2 cm at several clock sites in Germany and France [15], and with an uncertainty of about 3 cm in Italy [22], in UK [23], and in Canada [24]. Additionally, knowledge of the local geopotential is a mandatory criterion in the roadmap for redefining the second [12].

At Korea Research Institute of Standards and Science (KRISS), Yb optical lattice clocks [27-31] and Cs fountain clocks [32] have been developed. The first optical lattice clock at KRISS, KRISS-Yb1, has regularly been used since 2021 to calibrate the TAI [33, 34]. Thus, the heights of the atomic clocks at KRISS must be accurately estimated for steering TAI and improving clock comparison uncertainties.

In this study, we established methods for determining the height of an atomic clock. Surveying data required to apply each method were obtained, and the height of the atomic clock was determined based on these data. Differences between the determined heights were checked for consistency using various methods. In addition, the uncertainty of the determined height was estimated by considering the uncertainty of each surveying measurement. The remainder of this paper is organized as follows. The theoretical background is presented in section 2. In section 3, detailed data collection using the Global Navigation Satellite System (GNSS), spirit leveling, gravity surveying, and geopotential models are presented. Finally, the height of the atomic clock and its uncertainty are discussed in section 4.

## 2. Theory

In traditional geodesy, the height at point $P$ ($H_P$) is determined by dividing the geopotential number ($C_P$) by the average Helmert gravity at $P$ ($\bar{g}_P$), and $C_P$ is calculated as the difference between the reference gravity potential value on the geoid ($W_0$) and the gravity potential at $P$ ($W_P$) [35]:

$$H_P = \frac{C_P}{\bar{g}_P} = \frac{W_0 - W_P}{\bar{g}_P}. \qquad (2)$$

In modern geodesy, however, height is determined as the difference between the ellipsoidal height and the geoidal height rather than using the gravity potential (equation (3)) because the ellipsoidal height can be determined precisely and efficiently through GNSS surveying [35] as shown in Figure 1:

$$H_P = h_P - N_P, \qquad (3)$$

where $h_P$ is the ellipsoidal height of $P$, and $N_P$ is the geoidal height of $P$. $N_P$ is the difference between the geoid and the ellipsoidal surface.

Notably, the reference gravity potentials in equations (2) and (3) do not coincide with $W_0^{CGPM}$. If the reference gravity potential of a global geopotential model (GGM) is $W_0^{GGM}$ and a local geoid model (local vertical datum; LVD) is $W_0^{LVD}$, we should consider the biases as their differences from $W_0^{CGPM}$. Figure 1 shows the relationship between the reference surfaces and heights that should be considered in the computation.

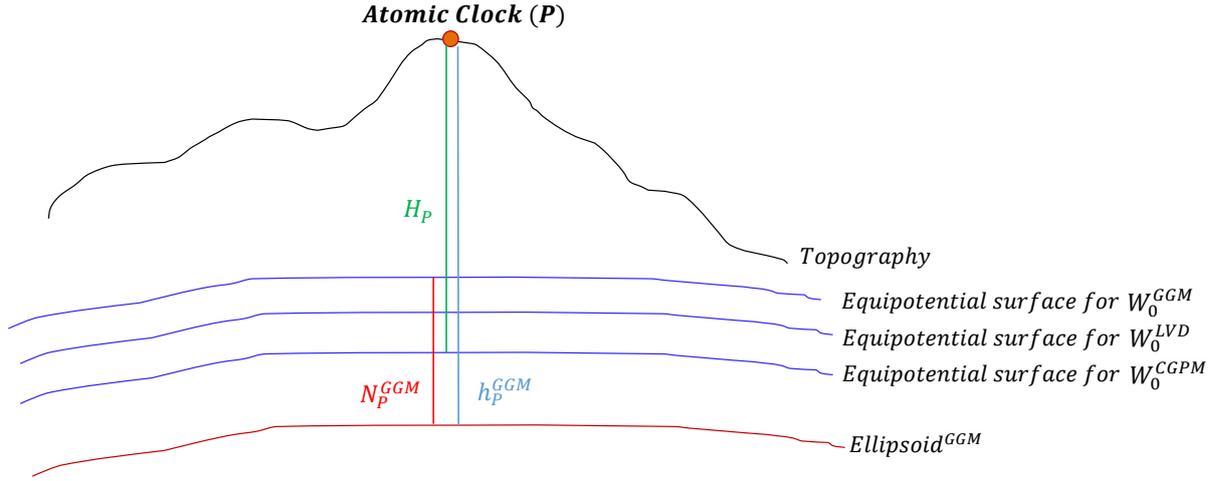

**Figure 1.** Reference surfaces and heights in this research. $H_P$: height at a point $P$ with respect to the equipotential surface for $W_0^{CGPM}$, $N_P^{GGM}$: geoidal height in a global geopotential model (GGM) at $P$, $h_P^{GGM}$: ellipsoidal height referenced in GGM, $W_0^{CGPM}$: reference potential conventionally adopted by CGPM, $W_0^{GGM}$: reference potential in GGM, $W_0^{LVD}$: reference potential in a local geoid model, $Ellipsoid^{GGM}$: reference ellipsoid in GGM.

Because the atomic clocks are in the basement of Building 313 in KRISS, GNSS surveying cannot be performed at the atomic clock location. Thus, an external control point (ECP) should be set for a GNSS surveying. The height referenced to $W_0^{CGPM}$ was determined at the ECP by considering bias correction between the reference surfaces. Next, the height difference between the ECP and the atomic clock is determined based on spirit leveling and gravity surveying, and the height of the atomic clock is determined by adding the height difference as follows:

$$H_Q = H_P + \Delta H_{PQ}, \qquad (4)$$

where $H_Q$ is the height of the atomic clock, $H_P$ is the height of the ECP referenced to $W_0^{CGPM}$, and $\Delta H_{PQ}$ is the height difference between the ECP ($P$) and the atomic clock position ($Q$) determined by spirit leveling and gravity surveying. Figure 2 briefly illustrates the method for determining the height of an atomic clock using equation (4).

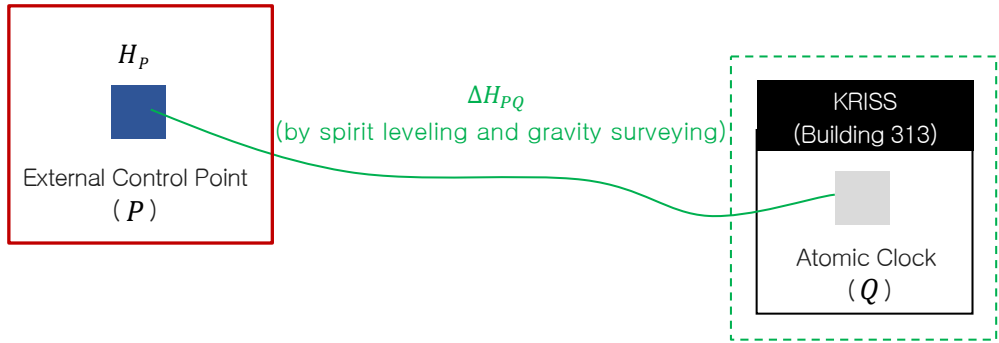

**Figure 2.** Concept used to determine the atomic clock height

### 2.1. Computation methods

Three computational methods were adopted, depending on whether the approach was traditional (equation (2)) or modern (equation (3)) or whether the geopotential model was global (GGM) or local.

**Method I:**

In Method I, the traditional approach (equation (2)) based on the GGM was applied. When applying a GGM, which is referenced to $W_0^{GGM}$, a correction factor ($\Delta H_{Correction}$) should be introduced that considers the difference from $W_0^{CGPM}$ as shown below.

$$H_P = \frac{C_P}{\bar{g}_P} + \Delta H_{Correction} = \frac{(W_0^{GGM} - W_P)}{\bar{g}_P} + \frac{(W_0^{CGPM} - W_0^{GGM})}{\bar{g}_P}. \qquad (5)$$

The term on the left in equation (5) represents the height from the reference surface of the GGM. The average Helmert gravity at $P$ ($\bar{g}_P$) was determined iteratively using the relation $\bar{g}_P = g_P + 0.0424 \times H_P$, where the units of $g_P$ and $H_P$ are mGal and m, respectively [36]. Because $H_P$ is required to calculate the $\bar{g}_P$, its initial value was calculated as the difference between the ellipsoidal height from the GNSS surveying and the geoidal height from the GGM.

**Method II:**

In Method II, the modern approach in equation (3) was applied based on the GGM. As in Method I, $\Delta H_{Correction}$ should be introduced because of the bias between $W_0^{GGM}$ and $W_0^{CGPM}$.

$$H_P = h_P - N_P^{GGM} + \Delta H_{Correction}, \qquad (6)$$

where $N_P^{GGM}$ is the geoidal height at $P$ in the GGM and $\Delta H_{Correction} = (W_0^{CGPM} - W_0^{GGM})/\bar{g}_P$.

**Method III:**

In this method, a modern approach was applied, as in Method II, but a local geoid model was used rather than using a GGM. Using this method, the applicability of a local model for computation can be verified and GGM-based methods can be validated. The geoidal height from the GGM ($N_P^{GGM}$) in equation (6) was changed to the local geoidal height $N_P^{LVD}$. Additionally, the bias term ($\Delta H_{Correction}$) in equation (6) was changed from $(W_0^{CGPM} - W_0^{GGM})/\bar{g}_P$ to $(W_0^{CGPM} - W_0^{LVD})/\bar{g}_P$. $W_0^{LVD}$ is typically calculated by adding the difference in gravity potential between the local vertical datum and GGM to the $W_0^{GGM}$. The gravity potential difference was determined by multiplying the normal gravity by the difference between the geoidal height from the GGM and the local geoidal height based on GNSS/leveling. Then, equation (6) is written as equation (7) by reflecting the use of the local geoid and bias term between $W_0^{LVD}$ and $W_0^{CGPM}$.

$$H_P = h_P - N_P^{LVD} + \Delta H_{Correction}^{LVD}$$
$$= h_P - N_P^{LVD} + (W_0^{CGPM} - W_0^{LVD})/\bar{g}_P$$
$$= h_P - N_P^{LVD} + \left\{W_0^{CGPM} - \left(W_0^{GGM} - \gamma_{avg} \times \frac{\sum_{i=1}^{k}(h_i - H_i - N_i^{GGM})}{k}\right)\right\}/\bar{g}_P, \quad (7)$$

where $h_i$ and $H_i$ are the ellipsoidal heights and heights of the locally distributed control points, respectively, $k$ is the number of the GNSS/leveling benchmarks, and $\gamma_{avg}$ is the average normal gravity on the reference ellipsoid over the test area. In principle, sufficient GNSS/leveling data with a homogeneous distribution and quality throughout the target area are required. Thus, $\Delta H_{Correction}^{LVD}$ in equation (7) was calculated using $k$ data available in total. Refer to [37] for detailed concepts for determining $W_0^{LVD}$.

In equations (5)–(7), a precise position determined by GNSS surveying is required to calculate the gravity potential and geoidal height, and gravity surveying is necessary to apply $\bar{g}_P$ when calculating the bias term.

### 2.2. Spirit leveling and orthometric correction

If the height of the ECP is determined, then the height of the atomic clock is determined by adding the height difference between the ECP and the atomic clock located inside the building, as shown in equation (4). The height difference in spirit leveling often varies depending on the surveying route because the leveling instrument is mounted in the plumb line direction; thus, gravity surveying should be performed to determine the unique value of the height difference. This type of correction ($OC_{PQ}$) is generally known as orthometric correction [35]. The height difference between the ECP and the atomic clock can be expressed using orthometric correction as follows [36]:

$$\Delta H_{PQ} = \delta n_{PQ} + OC_{PQ} = \delta n_{PQ} + \left(\sum_{i=P}^{Q}\left(\frac{g_i-\gamma_0}{\gamma_0}\right)\delta n_i + \left(\frac{\bar{g}_P-\gamma_0}{\gamma_0}\right)H_P - \left(\frac{\bar{g}_Q-\gamma_0}{\gamma_0}\right)H_Q\right), \quad (8)$$

where $\delta n_{PQ}$ is the height difference between $P$ and $Q$ which is determined by the spirit leveling only, $i$ is the leveling surveying route from $P$ to $Q$, $g_i$ is the gravity determined by gravity surveying, $\delta n_i$ is the measured height difference in the $i$-th interval, and $\gamma_0$ is normal gravity for an arbitrary standard latitude, typically 45° [36]. In equation (8), $P$ and $Q$ are the points of the ECP and the atomic clock, respectively, as in equation (4).

As shown in equation (8), $OC_{PQ}$ can be determined by obtaining the gravity at each point of the leveling route. The magnitude of $OC_{PQ}$ is usually smaller than the millimeter scale when the height difference between the two points is not large and the total leveling route is short. In this case, $OC_{PQ}$ is often dismissed, or only gravity is obtained at $P$ and $Q$ under the assumption that the height difference and gravity difference are linearly correlated. However, in this study, gravity data were obtained at all points of the leveling route for orthometric correction, with the concern that an abnormal force may be present because the atomic clock is in the basement of a building that is a massive concrete structure.

### 3. Data collection for height determination

We established the ECP in front of Building 313 at the KRISS, where the atomic clock is located. GNSS surveying and spirit leveling were conducted in cooperation with the National Geographic Information Institute (NGII) to obtain data for height determination [38]. Gravity surveying was performed by KRISS Superconducting Gravity Team for the orthometric correction. Figure 3 shows a schematic diagram of the surveying methods used in this study, including the surveying points and route.

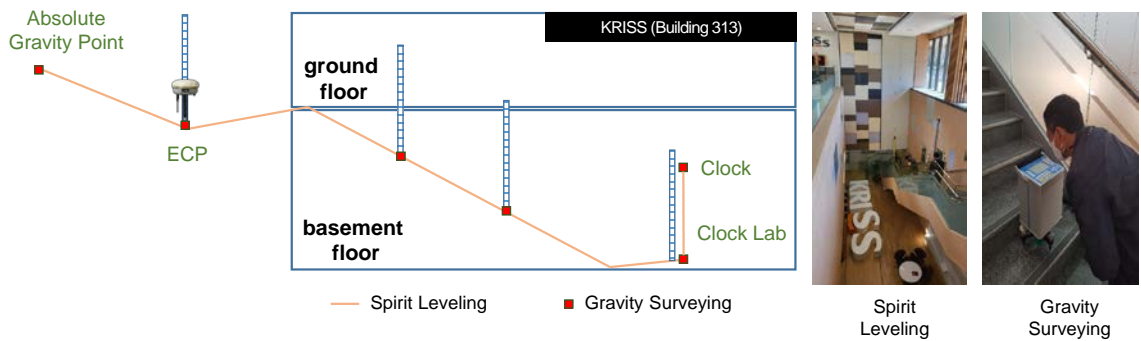

**Figure 3.** Schematic diagram of Global Navigation Satellite System (GNSS), spirit leveling, and gravity surveying. The surveying points and route are also shown.

### 3.1. GNSS surveying

Inside KRISS, four points (national control point UDJ51 and its complement points UDJ51A, KRISSB, and KRISSA) were established and will be maintained permanently to provide a precise three-dimensional position, gravity, and azimuth for various scientific studies. We selected KRISSA as the ECP because it was right in front of Building 313, where the atomic clock is located. To determine the precise position of the ECP as well as the other three points established in KRISS, GNSS data were obtained for 8 h on October 25, 2022. We conducted GNSS surveying at five national control points near KRISS (red dots in Figure 4), and four points in KRISS (blue dots in Figure 4) including the ECP. Relative positioning was performed to improve the reliability of surveying. Because the position of the unknown point is determined relative to one or more points with known position(s), GNSS data were obtained at five national control points that surround the newly established points at the same time to guarantee the stability of geometric distribution. The maximum baseline between the national control points and ECP was approximately 4.1 km. Figure 4 shows the distribution of the GNSS surveying points.

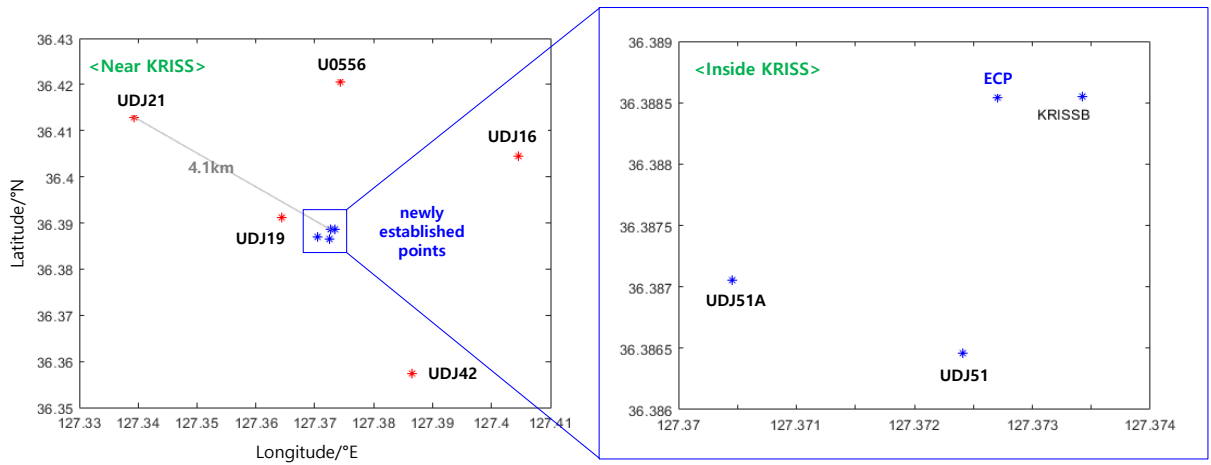

**Figure 4.** Distribution of the Global Navigation Satellite System (GNSS) surveying points

Three-dimensional positions of the ECP, KRISSB, UDJ51, and UDJ51A were determined using Bernese, Trimble Business Center (TBC), and Leica Geo Office (LGO) software. In relative positioning, the position of an undetermined point is determined by fixing the positions of the control points (baseline processing). Thus, the positions of the newly established points in KRISS were determined by fixing the officially announced positions of the five national control points. Three different types of software were used to confirm the reliability of the GNSS measurements by monitoring differences among the results of each software program, which differ in the detailed mechanism and modeling applied for data processing. The latitude, longitude, and ellipsoidal height of the ECP were 36.388537°N, 127.372709°E, and 106.284 m, respectively. The standard deviation of the results was assigned to the uncertainty of the ellipsoidal height $u_{h_P} = 0.022$ m.

### 3.2. Spirit leveling and gravity surveying

Spirit leveling and gravity surveying were conducted to determine the height difference between the ECP and the atomic clock. Spirit leveling was performed on October 26, 2022. Leveling was performed by setting the ECP to both the start and end points to ensure a round trip. Several points on the ground floor, basement floor, and clock lab in Building 313 were located on the leveling route. Round-trip leveling revealed a height difference of -0.00027 m during the round-trip, demonstrating the accuracy of the leveling.

Gravity surveying was also conducted at the ECP, at each of the surveying points along the leveling route, and at the bottom of the atomic clock lab to determine absolute gravity, which was designed using the absolute gravity point in KRISS as both the start and end points as shown in Figure 3. The gravity at each point was measured using a gravimeter (CG-5) with an uncertainty of 0.05 mGal. The absolute gravity at the atomic clock (specifically at the position of Yb atoms trapped in the optical lattice of KRISS-Yb1) was estimated to be 979 832.46(5) mGal using the generally known dependence of gravity at a height of 0.3086 mGal/m [35]. This was confirmed by using a relative gravimeter in December 2022.

Table 1 shows the height difference between the ECP and the atomic clock, which includes the orthometric correction ($OC_{PQ}$) using equation (8). In Table 1, $\delta n_{PQ}$ is the height difference determined only by spirit leveling, and $\Delta H_{corrected}$ is the height difference with correction by $OC_{PQ}$. $OC_{PQ}$ obtained by equation (8) was very small as shown in Table 1, and its uncertainty is discussed in Appendix A. Leveling and gravity surveying revealed that the atomic clock was 6.1469 m below the ECP. Because the orthometric correction was very small, the uncertainty of the height difference between the ECP and the atomic clock was given by the uncertainty of spirit leveling. This uncertainty $u_{\Delta H_{PQ}}$ was taken to be 0.0003 m, considering the height difference determined during the round-trip measurement of the spirit leveling.

**Table 1.** Height difference ($\Delta H_{PQ}$) between the ECP and atomic clock based on spirit leveling and gravity surveying (orthometric correction $OC_{PQ}$ is included)

| Leveling | | $\delta n_{PQ}$/m | Gravity/mGal | | $OC_{PQ}$/m | $\Delta H_{PQ}$/m |
|---|---|---|---|---|---|---|
| Start Point | End Point | | Start Point | End Point | | |
| ECP | Mid-Stair 1 | -0.955 5 | 979 831.07(5) | 979 831.20(5) | -0.000 004(6) | -0.955 5 |
| Mid-Stair 1 | Mid-Stair 2 | -2.244 9 | 979 831.20(5) | 979 831.64(5) | -0.000 021(6) | -2.245 0 |
| Mid-Stair 2 | Clock Lab | -4.069 0 | 979 831.64(5) | 979 832.76(5) | -0.000 060(5) | -4.069 1 |
| Clock Lab | Atomic Clock | 1.122 6 | 979 832.76(5) | 979 832.46(5) | 0.000 016(5) | 1.122 6 |
| **Total (ECP – Clock)** | | **-6.146 8(3)** | | | | **-6.146 9(3)** |

### 3.3. GGM

When applying equations (5) and (6) to determine the height of the ECP, a GGM is required to calculate the gravity potential or geoidal height of the ECP. In addition, $W_0^{GGM}$ should be calculated to match the reference surface to that recommended by CGPM [16]. Since the development of the first model (SE1) in 1966, more than 180 GGMs have been developed [39].

We applied EGM2008 [40], which was developed by the National Geospatial-Intelligence Agency (NGA) in 2009 and is one of the most widely used GGMs. We used the calculation service offered by International Center for Global Earth Models (ICGEM) [39]. The maximum degree and order for calculating the gravity potential and geoidal height were set to 2190 and 2159, and the reference ellipsoid was set to WGS84. Because the reference ellipsoid was set to WGS84, the reference gravity potential referred to by the GGM should be $W_0^{EGM2008(WGS84)}$. We determined the $W_0^{EGM2008(WGS84)}$ to be 62 636 851.719 m²s⁻² by the average of the gravity potential values at the geoid surface of the test area (11 km by 11 km) around KRISS with 1 arcmin grid. As discussed in [21], the height determination at a location depends only on the computed result of the (absolute) gravity potential, thus, the determined value of $W_0^{EGM2008(WGS84)}$ can be considered to be a constant with no uncertainty.

When we applied the position of the ECP determined in Section 3.1, the gravity potential and the geoidal height were calculated as 62 636 059.42 $m^2s^{-2}$ and 25.425 m, respectively. The uncertainty of the geoidal height calculated by the GGM was evaluated by comparing it with GNSS/leveling data in Korea. When the geoidal heights from EGM2008 were compared at 29 national control points within a radius of 10 km of the KRISS, the standard deviation of the differences was 0.030 m, which was considered as the uncertainty of the geoidal height calculated by EGM2008, $u_{N_P^{GGM}}$.

### 3.4. Local geopotential model and $W_0^{LVD}$

As shown in equation (7), Method III, which uses a local geoid model, was also used to confirm the validity of the GGM-based height using Method II and to check the applicability of the local geopotential model in height determination. We used a local geoid model, the Korean National Geoid 2018 (KNGeoid18) developed by NGII in 2018 [41]. The geoidal height was calculated as 25.254 m when KNGeoid18 was used. As in Method II using GGM, the difference between the reference surfaces, $W_0^{CGPM} - W_0^{LVD}$, should be considered in Method III. We determined $W_0^{LVD}$ to be 62 636 853.327 $m^2s^{-2}$ using equation (7) and the data at 29 national control points used in section 3.3. The uncertainty of $W_0^{LVD}$, $u_{W_0^{LVD}}$, was estimated to be 0.082 $m^2s^{-2}$ using equation (7), when we assume the uncertainties of the ellipsoidal height, the GGM geoidal height, and the orthometric height to be 0.030 m, 0.030 m and 0.015 m, respectively [42, 43].

### 4. Height determination at KRISS atomic clock

The height of the atomic clock was determined using the three methods described in section 2.1.

The process to obtain the result by Method I (equation (5)), which is a traditional approach based on the GGM, is shown in Table 2. The values of the components included in equation (5) and calculated heights of the ECP and the atomic clock are summarized. As described in section 3.3, the gravity potential of the external control point ($W_P$) was determined to be 62 636 059.42 m$^2$s$^{-2}$ with an uncertainty of 0.36 m$^2$s$^{-2}$, considering the uncertainty of the geoidal height (0.030 m) and that of the ellipsoidal height (0.022 m). The height of the ECP with respect to reference surface $W_0^{EGM2008(WGS84)}$ was 80.860(37) m. After correcting for bias between $W_0^{EGM2008(WGS84)}$ and $W_0^{CGPM}$, the height of the ECP was 81.297(37) m. By adding the height difference determined by spirit leveling and gravity surveying, the height of the atomic clock was calculated as 75.150(37) m.

**Table 2.** Height of the atomic clock determined using Method I (traditional method based on GGM)

| | Components | Value | Unit |
|---|---|---|---|
| ① | Gravity potential of the ECP ($W_P$) | 62 636 059.42(36) | m$^2$s$^{-2}$ |
| ② | Reference gravity potential referred by GGM ($W_0^{EGM2008(WGS84)}$) | 62 636 851.719 (constant) | m$^2$s$^{-2}$ |
| ③ | Reference gravity potential referred by CGPM ($W_0^{CGPM}$) | 62 636 856.000 (constant) | m$^2$s$^{-2}$ |
| ④ | Helmert gravity of the ECP ($\bar{g}_P$) | 9.798 345 0(5) | ms$^{-2}$ |
| ⑤ | Height of the ECP in terms of $W_0^{EGM2008(WGS84)}$ ($H_P = \frac{②-①}{④}$) | 80.860(37) | m |
| ⑥ | Bias between reference surfaces ($\Delta H_{correction} = \frac{③-②}{④}$) | 0.436 910 52(2) | m |
| ⑦ | Height of the ECP in terms of $W_0^{CGPM}$ ($H_P = ⑤ + ⑥$) | 81.297(37) | m |
| ⑧ | Height difference between the ECP and the atomic clock ($\Delta H_{PQ}$) | -6.146 9(3) | m |
| ⑨ | Height of the atomic clock ($H_Q = ⑦ + ⑧$) | **75.150(37)** | m |

In Method II (equation (6)), a modern approach based on the GGM was applied. Because the ellipsoidal and geoidal heights were computed as 106.284(22) and 25.425(30) m, respectively, the height of the ECP referenced to the surface of $W_0^{EGM2008(WGS84)}$ was 80.859(3) m. After applying the bias correction term and height difference as in Method I, the height of atomic clock was determined to be 75.149(37) m using Method II. The values of the components included in equation (6) and calculated heights of the ECP and the atomic clock are summarized in table 3.

**Table 3.** Height of the atomic clock determined using Method II (modern method based on GGM)

|   | Components | Value | Unit |
|---|---|---|---|
| ① | Ellipsoidal height of the ECP ($h_P$) | 106.284(22) | m |
| ② | Geoidal height of the ECP ($N_P$) | 25.425(30) | m |
| ③ | Height of the ECP in terms of $W_0^{EGM2008(WGS84)}$ ($H_P = ② - ①$) | 80.859(37) | m |
| ④ | Bias between reference surfaces ($\Delta H_{correction}$) | 0.436 910 52(2) | m |
| ⑤ | Height of the ECP in terms of $W_0^{CGPM}$ ($H_P = ③ + ④$) | 81.296(37) | m |
| ⑥ | Height difference between the ECP and the atomic clock ($\Delta H_{PQ}$) | -6.146 9(3) | m |
| ⑦ | Height of the atomic clock ($H_Q = ⑤ + ⑥$) | **75.149(37)** | m |

In Method III (equation (7)), a modern approach was applied based on a local geoid model, which is summarized in Table 4. As described in section 3.4, $W_0^{LVD}$ was determined to be 62 636 853.327(82) m²s⁻². $W_0^{CGPM}$ is considered to be a constant. The value of $h_P$ and its uncertainty are taken from section 3.1. $N_P^{LVD}$ was estimated 25.254 m, and its uncertainty is 0.023 m for KNGeoid18 geoid model [41]. The uncertainty of $N_P^{LVD}$ is relatively smaller than that of GGM because the local data with high precision and resolution were additionally used in the local geoid model. The height of the ECP referenced to $W_0^{LVD}$ and that referenced to $W_0^{CGPM}$ are estimated to be 81.030(32) m and 81.303(33) m, respectively. Using the height difference between the ECP and the atomic clock, the height of the atomic clock is determined to be 75.156(33).

**Table 4.** Height of the atomic clock determined using Method III (modern method based on LVD)

|  | Components | Value | Unit |
|---|---|---|---|
| ① | Reference gravity potential referred by LVD ($W_0^{LVD}$) | 62 636 853.327(82) | m²s⁻² |
| ② | Reference gravity potential referred by CGPM ($W_0^{CGPM}$) | 62 636 856.000 (constant) | m²s⁻² |
| ③ | Ellipsoidal height of the ECP ($h_P$) | 106.284(22) | m |
| ④ | Geoidal height of the ECP ($N_P^{LVD}$) | 25.254(23) | m |
| ⑤ | Helmert gravity of the ECP ($\bar{g}_P$) | 9.798 345 0(5) | ms⁻² |
| ⑥ | Height of the ECP in terms of $W_0^{LVD}$ ($H_P = ③ - ④$) | 81.030(32) | m |
| ⑦ | Bias between reference surfaces ($\Delta H_{correction}^{LVD} = \frac{②-①}{⑤}$) | 0.272 8(84) | m |
| ⑧ | Height of the ECP in terms of $W_0^{CGPM}$ ($H_P = ⑥ + ⑦$) | 81.303(33) | m |
| ⑨ | Height difference between the ECP and the atomic clock ($\Delta H_{PQ}$) | -6.146 9(3) | m |
| ⑩ | Height of the atomic clock ($H_Q = ⑧ + ⑨$) | **75.156(33)** | m |

## 5. Conclusion

We applied three methods to determine the height of the atomic clock (KRISS-Yb1) developed at KRISS, utilizing data obtained by GNSS surveying, leveling surveying, gravity surveying, GGM, and local geoid models. The heights of the atomic clock calculated using Methods I, II, and III were 75.150(37), 75.149(37), and 75.156(33) m, respectively. The results of the three methods agree well within their uncertainties. Finally, the height of the atomic clock was determined to be 75.15 m by calculating the weighted mean of the three results. Because three results are not statistically independent, and have strong correlation among them, the uncertainty of the weighted mean was taken to be 0.04 m to be conservative. The relativistic redshift of KRISS-Yb1 was evaluated to be $8.193(4) \times 10^{-15}$ by equation (1).

**Appendix A. Uncertainty estimation of orthometric correction**

In this research, we applied orthometric correction in spirit leveling, though this correction was expected to be very small when the height difference was not large and the total leveling route was short. This was done because we wanted to confirm that the effect of a massive concrete structure of the building, where the atomic clock was located, was ignorable, because it was not self-evident. The results in Table 1 shows that the orthometric correction was ignorable compared to the uncertainty of the spirit leveling measurement.

In this section, we estimate the uncertainty of the orthometric correction. The orthometric correction was applied to each step of the surveying. If we assume the start point and the end point of a step to be $A$ and $B$ in equation (8), the orthometric correction in this step is given by

$$OC_{AB} = \frac{1}{\gamma_0} \left[ \left(\frac{g_A + g_B}{2} - \gamma_0\right)(H_B - H_A) + (\bar{g}_A - \gamma_0)H_A - (\bar{g}_B - \gamma_0)H_B \right]. \qquad (9)$$

We assume the uncertainties of the gravity measurement at $A$ and $B$ are $u(g_A)$ and $u(g_B)$, respectively, and $H_A$ and $H_B$ are given by spirit leveling. Then, the uncertainty of $OC_{AB}$ can be written as

$$u(OC_{AB}) = \sqrt{\left(\frac{\partial (OC_{AB})}{\partial g_A} u(g_A)\right)^2 + \left(\frac{\partial (OC_{AB})}{\partial g_B} u(g_B)\right)^2} \approx \frac{u(g)}{\sqrt{2}\gamma_0}(H_A + H_B), \qquad (10)$$

where $u(g) = u(g_A) = u(g_A)$ was 0.05 mGal. As a result, the uncertainty of $OC_{AB}$ in each step in Table 1 was estimated to be 0.000 005 m ~ 0.000 006 m.